\documentclass[aps,onecolumn,amsmath,amssymb,superscriptaddress,nofootinbib,11pt,a4paper]{revtex4}
\usepackage{titlesec}
\titleformat{\section}
  {\normalfont\Large\bfseries}  
  {\thesection}                 
  {1em}                         
  {}                            
\titleformat{\subsection}
  {\normalfont\large\bfseries}
  {\thesubsection}
  {1em}
  {}

\usepackage{float}
\usepackage{orcidlink}
\usepackage{multirow}

\begin{document} 

\title{Energy extraction via magnetic reconnection from rotating regular black holes in nonminimally coupled Einstein-Yang-Mills theory}

\author{Gao-Ming Deng}
\email{gmdeng2010@126.com}
\affiliation{ School of Physics and Astronomy, China West Normal University, Sichuan 637002, China}

\author{Si-Yuan Pan}
\email{pansiyuanphys@outlook.com}
\affiliation{ School of Physics and Astronomy, China West Normal University, Sichuan 637002, China}

\author{San-Duo Liu}
\affiliation{ School of Physics and Astronomy, China West Normal University, Sichuan 637002, China}

\author{Zong-Yao Zhang}
\affiliation{ School of Physics and Astronomy, China West Normal University, Sichuan 637002, China}

\begin{abstract}
We extend to investigate energy extraction via magnetic reconnection from rotating regular black holes in nonminimally coupled Einstein–Yang–Mills theory within the Comisso–Asenjo mechanism. Both circular orbit motion and plunging motion from the innermost stable circular orbit are considered. We analyze the effects of the magnetic charge $Q$, coupling parameter $\zeta$, magnetization parameter $\sigma$, and orientation angle $\xi$ on the outflow energy, allowed energy extraction regions, power, and efficiency. For the circular orbit region, the magnetic charge turns out to play different roles in the energy of the decelerated plasma for different reconnection radius. We define the critical boundary of energy $e_-^{\infty}$ evolution as the reversal boundary. Further analysis reveals that increasing $\zeta$ shifts this boundary toward larger radii, thereby broadening the radial range over which an increase in the magnetic charge reduces the energy of the decelerated plasma. Within the parameter range considered, increasing the coupling parameter causes only a slight inward shift of the outer boundary of the equatorial ergoregion, whereas the allowed energy extraction region in the circular orbit region shrinks considerably. In contrast, the plunging region retains a similar overall structure of the allowed energy extraction region over the common spin range. Furthermore, unlike the circular orbit region, the allowed energy extraction region in the plunging region does not expand monotonically with decreasing orientation angle $\xi$. Under the parameter configurations, the plunging region exhibits higher energy extraction power and efficiency. 
\end{abstract}
\maketitle
\newpage

\section{Introduction} \label{SecIntro}
Magnetic reconnection is an important physical process for the efficient release of magnetic energy in plasmas \cite{Sironi2014,Guo2014}. It has been widely invoked to explain various astrophysical phenomena, including solar flares \cite{Shibata2011} and coronal mass ejections \cite{Lin2000,Zweibel2009}. Current sheets and magnetic reconnection have also been investigated in numerical simulations of black-hole magnetospheres \cite{Parfrey2019,Bransgrove2021}. Comisso and Asenjo developed an analytical region for energy extraction through magnetic reconnection within the ergosphere of a Kerr black hole \cite{Comisso2021}. Under suitable conditions, one of the reconnection outflows carries negative energy at infinity and falls into the black hole, while the other carries positive energy and escapes, thereby enabling the net extraction of the black hole's rotational energy. Subsequently, this mechanism has been extended to various rotating black hole backgrounds. Some studies have considered only circular orbital motion \cite{Khodadi2022,Wei2022,Wang2022,LiRegular2023,LiHairy2023,ZhangKRZ2024,ZhangMagnetized2024,Shaymatov2024}, whereas others have investigated both circular and plunging motion \cite{Zeng2025,Yao2026,ZengKBR2025,WangZeng2025}. It has demonstrated that the energy extraction conditions, power, and efficiency are all influenced by the background geometry and the dynamical state of the plasma.

Rotating black holes in nonminimally coupled Einstein-Yang-Mills (EYM) theory provide a concrete setting for investigating the aforementioned effects. The black hole spacetime is characterized by the Wu-Yang magnetic charge $Q$ and the nonminimal coupling parameter $\zeta$, which modifies the event horizon, the outer boundary of the ergoregion, the photon circular orbits, and the maximum allowed spin. In nonminimal EYM theory, static spherically symmetric black hole solutions carrying Wu-Yang magnetic charge have been constructed, and their regularity and horizon structures have been investigated \cite{Balakin2016Regular,Balakin2016Exact}. Previous studies have also examined geodesic motion and Penrose energy extraction in this background \cite{Huo2026}. The energy extraction via magnetic reconnection also depends on the plasma magnetization, the outflow direction, and the state of motion. In particular, when the plasma plunges inward from the innermost stable circular orbit (ISCO), the effects of radial motion cannot be neglected, and its energy extraction properties may differ from those of the circular orbit region \cite{Zeng2025,Yao2026}. Therefore, it is necessary to further investigate whether and how the magnetic charge, the reconnection radius and the nonminimal coupling parameter affect the reversal boundary. Moreover, we also compare the energy extraction conditions and capabilities in the circular orbit region and the plunging region.

In this work, we consider both a circular orbit region and a plunging region in which the plasma moves inward from the ISCO. We investigate the effects of the magnetic charge $Q$, coupling parameter $\zeta$, magnetization parameter $\sigma$, and orientation angle $\xi$ on the outflow energy and the allowed energy extraction regions. For the circular orbit region, we further use the partial derivative of the energy at infinity per unit enthalpy of the decelerated plasma with respect to the magnetic charge as a diagnostic to quantitatively characterize the boundary at which the response to the magnetic charge reverses. We then examine how the nonminimal coupling parameter modifies this boundary, providing a complementary perspective for understanding the radial dependence of the magnetic-charge effect. Furthermore, we compare the differences in the allowed energy extraction regions and their dependence on the orientation angle between the two regions of plasma motion. 

The remainder of this paper is organized as follows. In Sec. \ref{SecEYMBH}, we introduce the background geometry of rotating regualr black holes in nonminimally coupled EYM theory. In Sec. \ref{SecEEC}, we investigate energy extraction via magnetic reconnection in the circular orbit region.  Section \ref{SecEEP} extends the analysis to the plunging region and compares the energy extraction capabilities of the two plasma configurations. Finally, Section \ref{SecCOn} summarizes our main results. For convenience, we set $G=c=1$ throughout this paper.

\section{Geometry of rotating regular black holes in nonminimally coupled Einstein–Yang–Mills theory} \label{SecEYMBH}

In Boyer-Lindquist coordinates, the line element of rotating regular black holes in nonminimally coupled EYM theory is given by \cite{Huo2026,Jusufi2021}
\begin{equation}
\begin{aligned}
ds^2=-\left[1-\frac{2m(r)r}{\rho^2}\right]dt^2+\frac{\rho^2}{\Delta}\,dr^2+\rho^2\,d\theta^2+\frac{\Sigma\sin^2\theta}{\rho^2}\,d\phi^2-\frac{4am(r)r\sin^2\theta}{\rho^2}\,dt\,d\phi ,
\end{aligned}
\label{eq:eymdg}
\end{equation}
and
\begin{equation}
\begin{aligned}
&\rho^2 = r^2+a^2\cos^2\theta,\\
&\Delta = r^2-2m(r)r+a^2,\\
&\Sigma = (r^2+a^2)^2-a^2\Delta\sin^2\theta,\\
&m(r) = \frac{r^3(2M\,r-Q^2)}{2(r^4+\lambda)},
\end{aligned}
\label{eq:eym_metric_functions}
\end{equation}
where $M$ denotes the black hole mass, $a=J/M$ is the spin parameter, $J$ is the angular momentum of black holes, $Q$ is the magnetic charge associated with the Wu-Yang gauge field, and $\zeta>0$ is the parameter characterizing the nonminimal coupling between the Yang-Mills field and the gravitational field with $\lambda=2\zeta Q^2$. In the limit $Q=0$, we have $\lambda=0$ and $m(r)=M$, and the line element above reduces to the Kerr black hole case.

The horizons of black holes are determined by $\Delta(r)=r^2-2rm(r)+a^2=0$, with the event horizon corresponding to the largest positive root of this equation. The boundary of the ergoregion is determined by $g_{tt}=0$. The region between the event horizon and the ergoregion boundary constitutes the ergoregion.

We consider particle motion in the equatorial plane ($\theta=\pi/2$) of the rotating EYM black hole and set $M=1$. The radial geodesic equation can be written as
\begin{equation}
    r^4\left(\frac{dr}{d\tau}\right)^2=[E(r^2+a^2)-aL]^2-\Delta(r)\left[\varepsilon r^2+(L-aE)^2\right]=\mathcal R(r),
    \label{eq:jxs}
\end{equation}
where $\mathcal R(r)$ is the radial effective potential governing the radial motion of the particle, $E$ and $L$ are the energy and angular momentum of the particle respectively \cite{Hod:2011aa,Deng:2014bta,Deng:2017bjz,Deng:2017yse}, and $\tau$ denotes the proper time. For photons $\varepsilon=0$, whereas $\varepsilon=1$ is for massive particles. To determine the radii of equatorial circular photon orbits, we set $\varepsilon=0$ and impose the circular orbit conditions \cite{Bardeen1972}
\begin{equation}
\mathcal R(r)=0, \qquad
\frac{\partial\mathcal R}{\partial r}=0.
    \label{eq:gzqbj}
\end{equation}
The equatorial circular photon orbits can be classified into prograde and retrograde orbits, corresponding to motion in the same and opposite directions as the black hole rotation respectively.

\begin{figure}[H]
    \centering
    \includegraphics[
        width=\linewidth,
        height=0.45\textheight,
        keepaspectratio
    ]{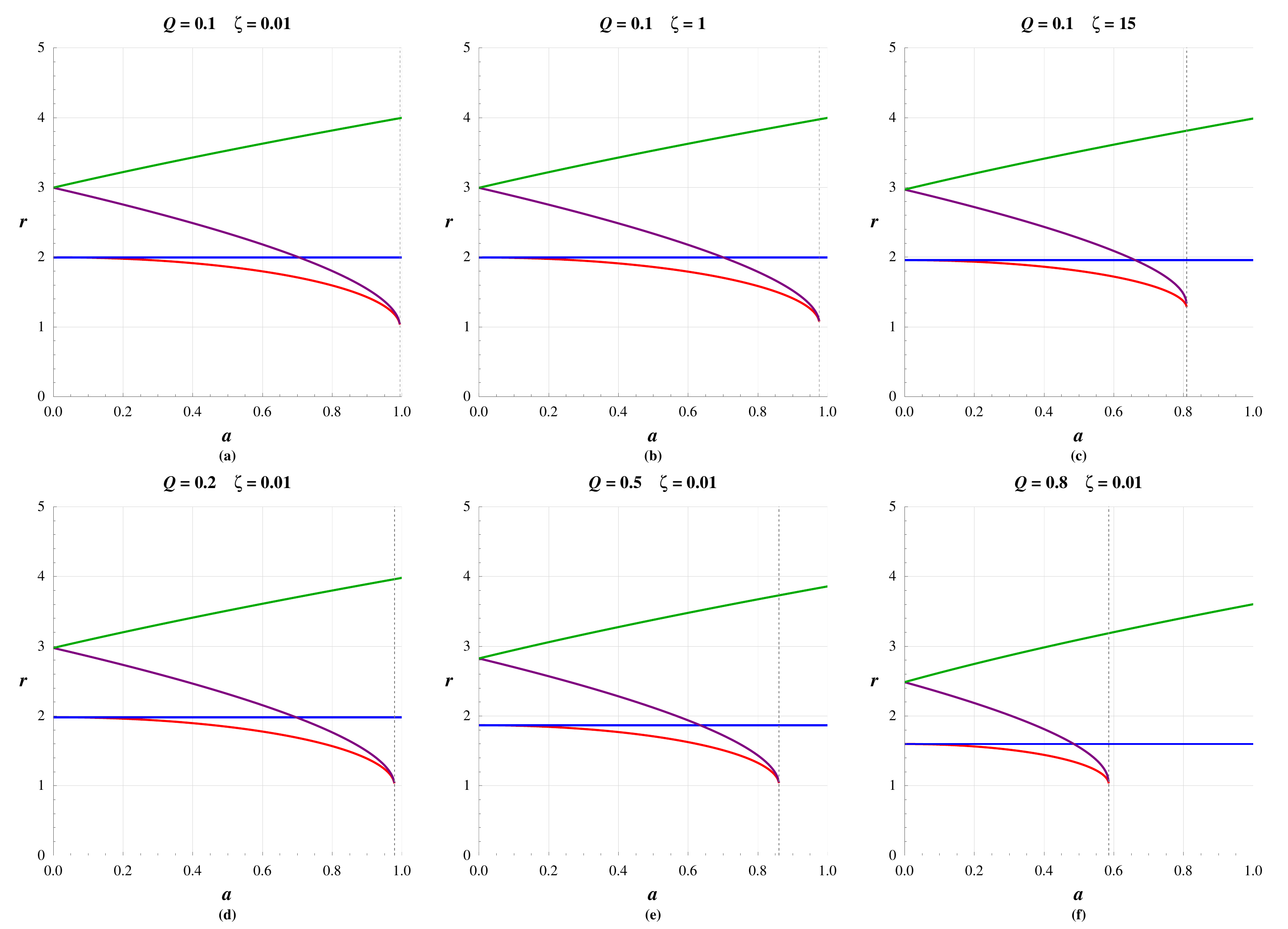}
    \caption{(Color online) Variation of the event horizon radius, ergoregion boundary, and photon orbit radii with the spin parameter $a$.}
    \label{fig:4r}
\end{figure}

Fig.~\ref{fig:4r} shows the variations of the event horizon, ergoregion boundary, and prograde and retrograde circular photon orbits with the spin parameter $a$ for different values of the magnetic charge $Q$ and coupling parameter $\zeta$. The red, blue, purple, and green solid curves represent the four boundaries described above respectively, while the gray vertical dashed lines indicate the maximum allowed spin for the corresponding backgrounds. For fixed $Q$ and $\zeta$, as the spin increases, the event horizon and the prograde circular photon orbit move toward smaller radii, whereas the retrograde circular photon orbit moves toward larger radii. The ergoregion boundary remains unchanged, leading to an increasing radial separation between the event horizon and the outer boundary of the ergoregion.

A comparison of Figs.~\ref{fig:4r}(a)-(c) shows that increasing the coupling parameter primarily leads to a decrease in the maximum allowed spin, while the outer boundary of the ergoregion shifts only slightly inward. A further comparison of Fig.~\ref{fig:4r}(a) with Figs.~\ref{fig:4r}(d)-(f) reveals that, as the magnetic charge increases, the outer boundary of the ergoregion moves noticeably toward smaller radii. Within the common spin range, the radii of the event horizon and both photon circular orbits also decrease, accompanied by a reduction in the maximum allowed spin of the black hole.

\section{Energy Extraction in Circular Orbits} \label{SecEEC}
\subsection{Magnetic Reconnection in Circular Orbits}

To describe the local plasma motion near the rotating EYM black hole, we adopt the reference frame of the zero-angular-momentum observer (ZAMO) \cite{Bardeen1972,Comisso2021}. In this frame, the line element can be expressed as
\begin{equation}
ds^2=-d\hat t^{\,2}+\sum_{i=1}^{3}(d\hat x^i)^2,
    \label{eq:zamo}
\end{equation}
where
\begin{equation}
d\hat t=\alpha\,dt,\qquad
d\hat x^i=\sqrt{g_{ii}}\,dx^i-\alpha\beta^i\,dt,
    \label{eq:zamofl}
\end{equation}
with the shift vector given by $\beta^i=(0,0,\beta^\phi)$, where no summation over $i$ is implied in the above expression. We also have
\begin{equation}
\alpha=\sqrt{-g_{tt}+\frac{g_{t\phi}^2}{g_{\phi\phi}}},
\qquad
\omega^\phi=-\frac{g_{t\phi}}{g_{\phi\phi}},
\qquad
\beta^\phi=\frac{\sqrt{g_{\phi\phi}}\,\omega^\phi}{\alpha},
    \label{eq:zamogeom}
\end{equation}
where $\alpha$ is the lapse function, $\omega^\phi$ is the angular velocity of frame dragging.

For prograde circular orbits in the equatorial plane, the Keplerian angular velocity \cite{Harko2009,Huo2026} is given by
\begin{equation}
\Omega_K=
\frac{
-\partial_r g_{t\phi}
+\sqrt{(\partial_r g_{t\phi})^2
-(\partial_r g_{tt})(\partial_r g_{\phi\phi})}
}{
\partial_r g_{\phi\phi}}.
\label{eq:kpljsd}
\end{equation}

The background plasma velocity and its Lorentz factor as measured by a ZAMO are respectively given by
\begin{equation}
\hat v_K
=\frac{\sqrt{g_{\phi\phi}}}{\alpha}
(\Omega_K-\omega^\phi),
\qquad
\hat\gamma_K=(1-\hat v_K^2)^{-1/2}.
    \label{eq:sdyz}
\end{equation}

Within the Comisso-Asenjo magnetic reconnection region, under the assumptions of efficient magnetic energy conversion, adiabaticity, and incompressibility, the energy at infinity per unit enthalpy of the plasma can be written as
\begin{equation}
\begin{aligned}
e_\pm^\infty=\alpha\hat\gamma_K\Bigg[(1+\beta^\phi\hat v_K)\sqrt{1+\sigma}
\pm\cos\xi\,(\beta^\phi+\hat v_K)\sqrt{\sigma}-\frac{\sqrt{1+\sigma}\mp\cos\xi\,\hat v_K\sqrt{\sigma}}{4\hat\gamma_K^2\left(1+\sigma-\cos^2\xi\,\hat v_K^2\sigma\right)}\Bigg]，
\end{aligned}
    \label{eq:wcqnl}
\end{equation}
where $\sigma=B_0^2/w_0$ is the magnetization parameter of the upstream plasma in the reconnection layer, where $B_0$ and $w_0$ denote the magnetic field strength and enthalpy density in the upstream comoving frame respectively. The angle $\xi$ is defined as the angle between the outflow direction and the azimuthal direction in the plasma comoving frame. The symbol $``\pm"$corresponds to the accelerated and decelerated outflows respectively.

We assume a relativistically hot plasma with an adiabatic index $\Gamma=4/3$ and $w=4p$. Similar to the original Penrose process \cite{Penrose1969}, the conditions for energy extraction from the black hole via magnetic reconnection are
\begin{equation}
e_-^\infty<0
,\qquad
\Delta e_+^\infty=e_+^\infty
-\left(1-\frac{\Gamma}{\Gamma-1}\frac{p}{w}\right)
=e_+^\infty>0.
    \label{eq:nltqtj}
\end{equation}
When the decelerated plasma carries negative energy at infinity, while the accelerated plasma acquires positive energy at infinity, the rotational energy of the black hole can be extracted.

\begin{figure}[H]
    \centering
    \includegraphics[
        width=\linewidth,
        height=0.45\textheight,
        keepaspectratio
    ]{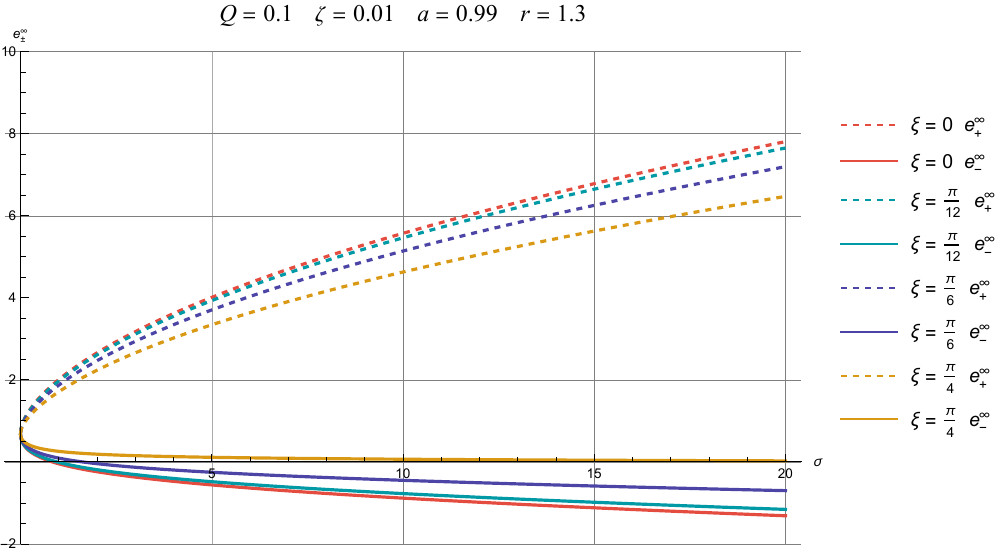}
    \caption{(Color online) Variation of the energy $e_{\pm}^{\infty}$ at infinity per unit
    enthalpy with the plasma magnetization parameter $\sigma$ for different orientation angle $\xi$.}
    \label{fig:e+e-xi}
\end{figure}

Fig.~\ref{fig:e+e-xi} shows the variation of the energy at infinity per unit enthalpy with the magnetization parameter for different orientation angles. Within the parameter range considered, $e_+^{\infty}$ remains positive throughout, so the key condition for energy extraction is whether $e_-^{\infty}$ becomes negative. As the plasma magnetization parameter $\sigma$ increases, $e_+^{\infty}$ increases, whereas $e_-^{\infty}$ generally decreases, thereby supporting energy extraction. At a fixed magnetization parameter, the increasing orientation angle $\xi$ decreases $e_+^{\infty}$ and increases $e_-^{\infty}$, thereby making energy extraction more difficult. A smaller orientation angle $\xi$ corresponds to a larger azimuthal velocity component of the outflow, making it easier for the decelerated plasma to enter the negative-energy regime and thus supporting the extraction of the black hole's rotational energy through magnetic reconnection. These are consistent with previous studies in Kerr and Kerr-Sen-AdS$_4$ backgrounds\cite{Comisso2021,Zeng2025}.

\begin{figure}[H]
    \centering
    \includegraphics[
        width=\linewidth,
        height=0.45\textheight,
        keepaspectratio
    ]{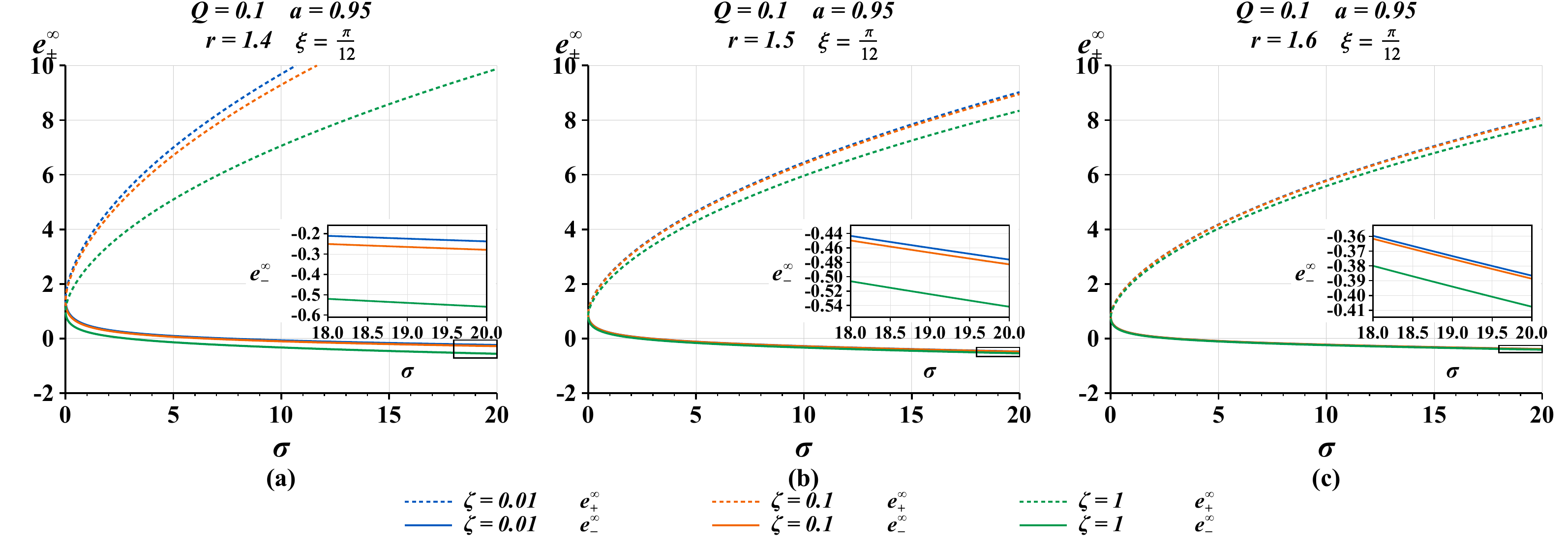}
    \caption{(Color online) Variation of the energy $e_{\pm}^{\infty}$ at infinity per unit enthalpy with the plasma magnetization parameter $\sigma$ for different reconnection radius $r$ and coupling parameter $\zeta$.}
    \label{fig:e+e-zeta}
\end{figure}

Fig.~\ref{fig:e+e-zeta} shows the variation of $e_{\pm}^{\infty}$ with $\sigma$ for different values of the coupling parameter $\zeta$. As $\sigma$ increases, $e_+^{\infty}$ continues to increase, whereas $e_-^{\infty}$ decreases and enters the negative-energy regime, indicating that a higher degree of magnetization favors the satisfaction of the energy extraction condition for magnetic reconnection. Furthermore, as $\zeta$ increases, $e_-^{\infty}$ attains more negative values of the energy at infinity. This effect is particularly pronounced at $r=1.4$, but becomes weaker at larger radii. In other words, as $r$ increases, the differences in $e_-^{\infty}$ among different values of $\zeta$ gradually decrease.

\begin{figure}[H]
    \centering
    \includegraphics[
        width=\linewidth,
        height=0.45\textheight,
        keepaspectratio
    ]{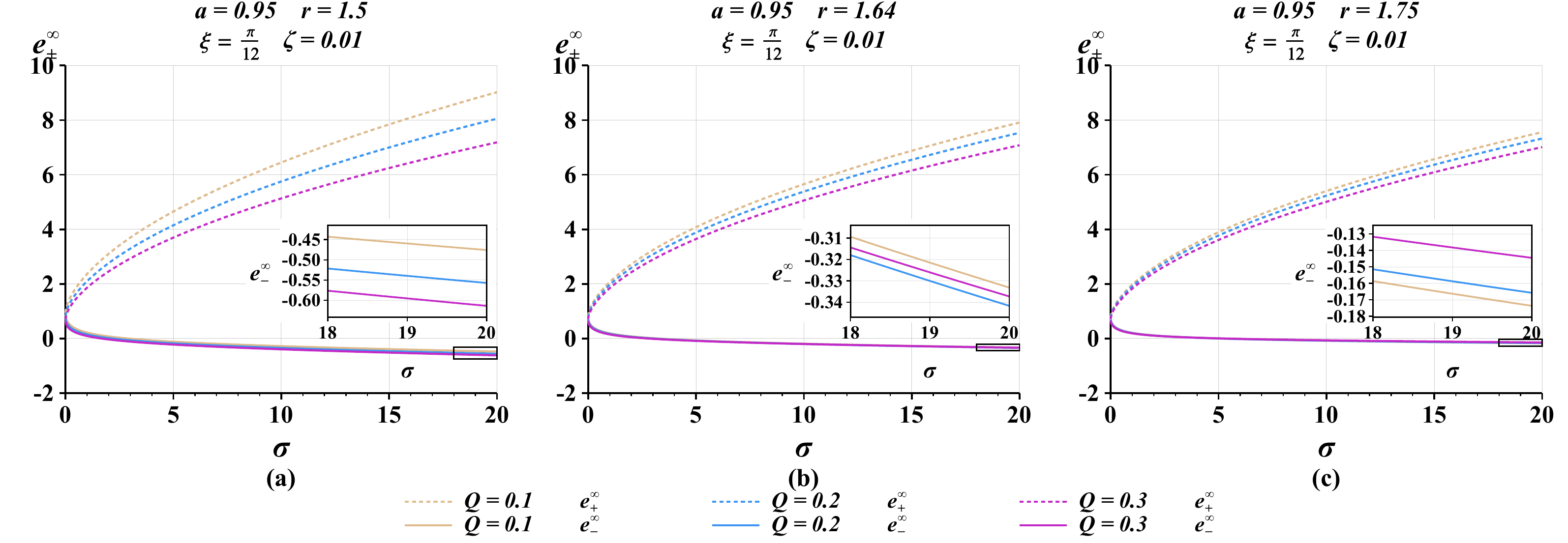}
    \caption{(Color online) Variation of the energy $e_{\pm}^{\infty}$ at infinity per unit
    enthalpy with the plasma magnetization parameter $\sigma$ for different reconnection radius $r$ and magnetic charge $Q$.}
    \label{fig:e+e-q}
\end{figure}

\begin{figure}[H]
    \centering
    \includegraphics[
        width=\linewidth,
        height=0.45\textheight,
        keepaspectratio
    ]{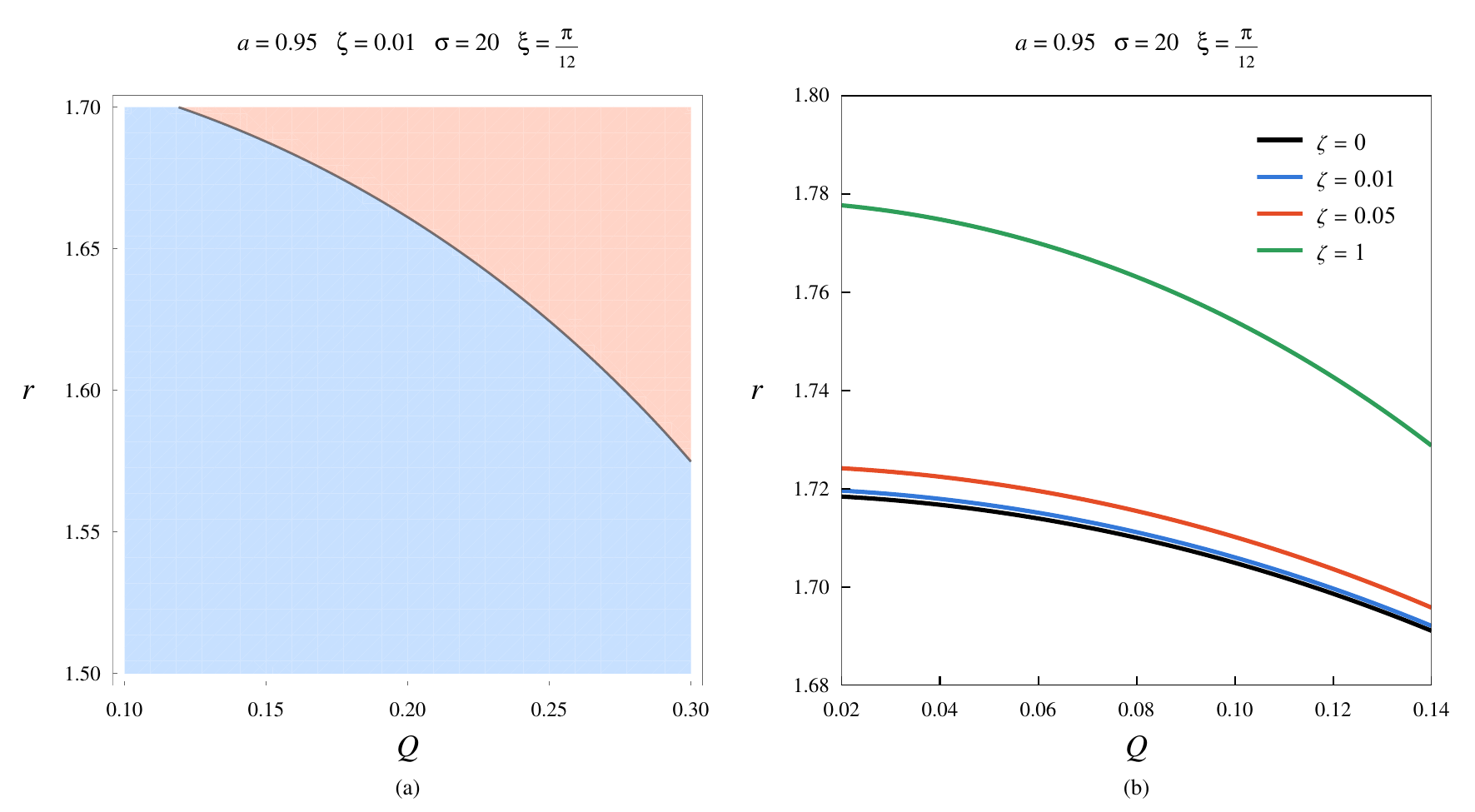}
    \caption{(Color online) Reversal boundary describing the dependence of $e_-^{\infty}$ on the magnetic charge $Q$ and the reconnection radius $r$.}
    \label{fig:qre-}
\end{figure}

Fig.~\ref{fig:e+e-q} shows the variation of $e_{\pm}^{\infty}$ with $\sigma$ for different values of the magnetic charge $Q$. As $\sigma$ increases, $e_+^{\infty}$ continues to increase, whereas $e_-^{\infty}$ decreases and enters the negative-energy regime. At fixed $r$ and $\sigma$, increasing $Q$ reduces the energy of the accelerated outflow  $e_+^{\infty}$, whereas the response of $e_-^{\infty}$ to the magnetic charge depends strongly on the magnetic reconnection radius $r$. Observing the closeup from the left subfigure for $r=1.5$, the energy $e_-^{\infty}$ decreases with increasing $Q$, whereas the opposite trend occurs for $r=1.75$ in the right subfigure. For the intermediate radius $r=1.64$, $e_-^{\infty}$ varies with the magnetic charge nonmonotonically. To further determine the location at which this trend reverses, Fig.~\ref{fig:qre-} examines the monotonic dependence of $e_-^{\infty}$ on $Q$ in the $(Q,r)$ plane at a fixed magnetization parameter $\sigma=20$. For convenience, the critical boundary of energy $e_-^{\infty}$ varying with the magnetic charge $Q$ is defined as reversal boundary, which can be determined by
\begin{equation}
    \frac{\partial e_-^\infty}{\partial Q}=0.
    \label{eq:qe-bj}
\end{equation}

Fig.~\ref{fig:qre-}(a) shows the reversal boundary, represented by the black solid curve. In the blue region below the boundary, increasing magnetic charge $Q$ decreases $e_-^{\infty}$. In contrast, in the red region above the boundary, increasing $Q$ increases the energy of the decelerated outflow. This division explains the changes in the ordering of $e_-^{\infty}$ at different radii observed in Fig.~\ref{fig:e+e-q}. The reversal boundary moves toward smaller radii as $Q$ increases, indicating that the location of the reversal also depends on the magnetic charge itself. Fig.~\ref{fig:qre-}(b) further compares the reversal boundaries for different coupling parameters. Within the parameter range shown, the reversal boundary shifts toward larger radii as the coupling parameter $\zeta$ increases, indicating that the nonminimal coupling can regulate the location of the reversal. For a fixed magnetic charge $Q$, a larger $\zeta$ broadens the radial range over which increasing the magnetic charge reduces $e_-^{\infty}$.

\subsection{Parameter Space for Allowed Energy Extraction via Magnetic Reconnection in Circular Orbits}

\begin{figure}[H]
    \centering
    \includegraphics[
        width=\linewidth,
        height=0.45\textheight,
        keepaspectratio
    ]{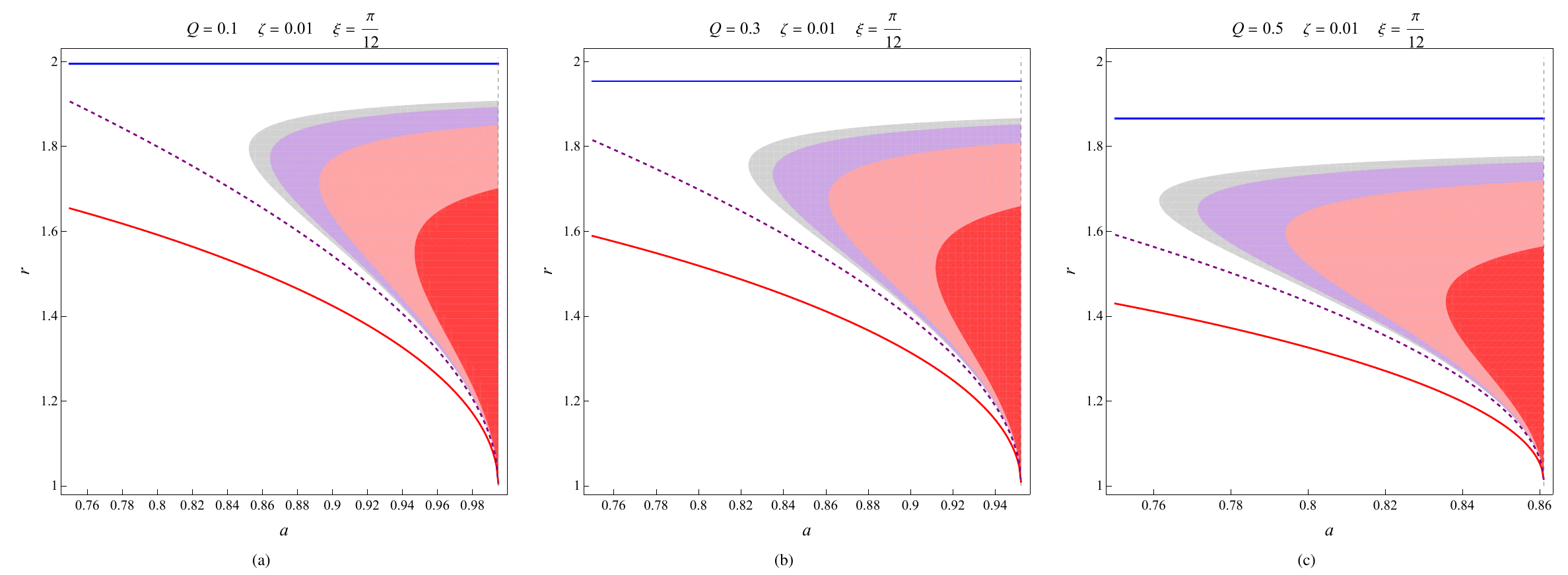}
    \caption{(Color online) Allowed energy extraction regions for different
    magnetic charges $Q$.}
    \label{fig:xyqqnltq}
\end{figure}

\begin{figure}[H]
    \centering
    \includegraphics[
        width=\linewidth,
        height=0.45\textheight,
        keepaspectratio
    ]{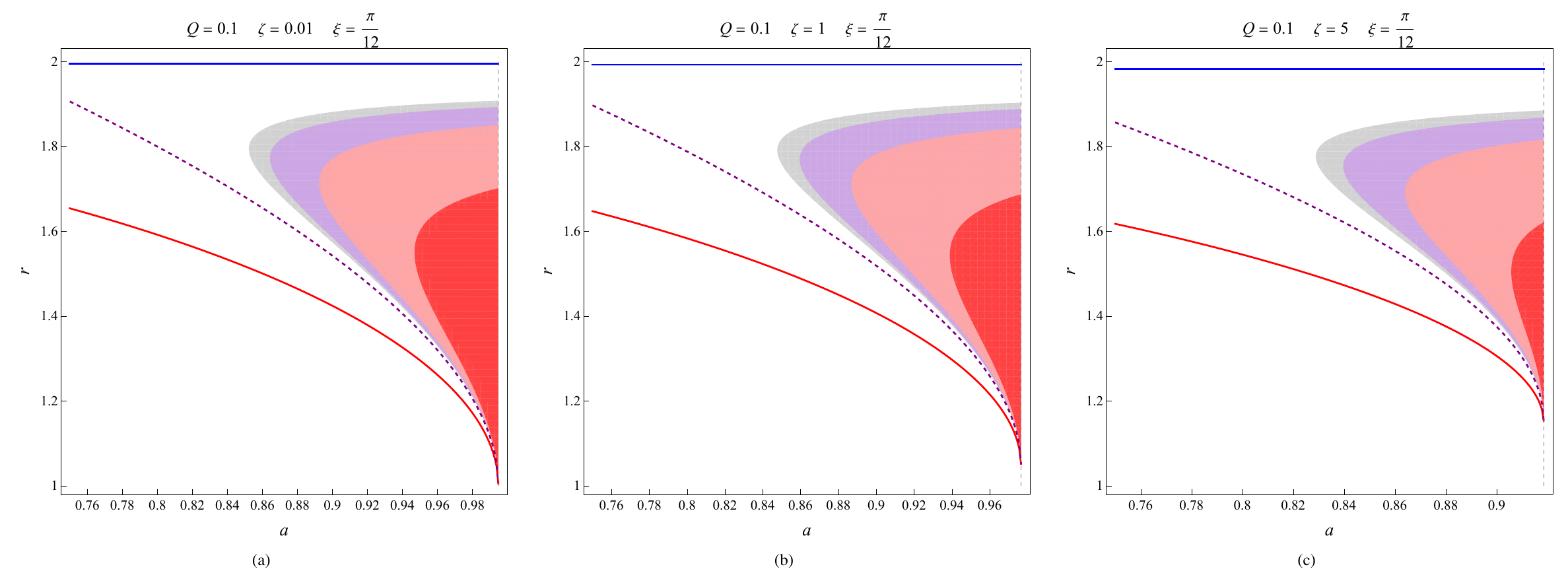}
    \caption{(Color online) Allowed energy extraction regions for different
    coupling parameters $\zeta$.}
    \label{fig:yxqzetalntq}
\end{figure}

\begin{figure}[H]
    \centering
    \includegraphics[
        width=\linewidth,
        height=0.45\textheight,
        keepaspectratio
    ]{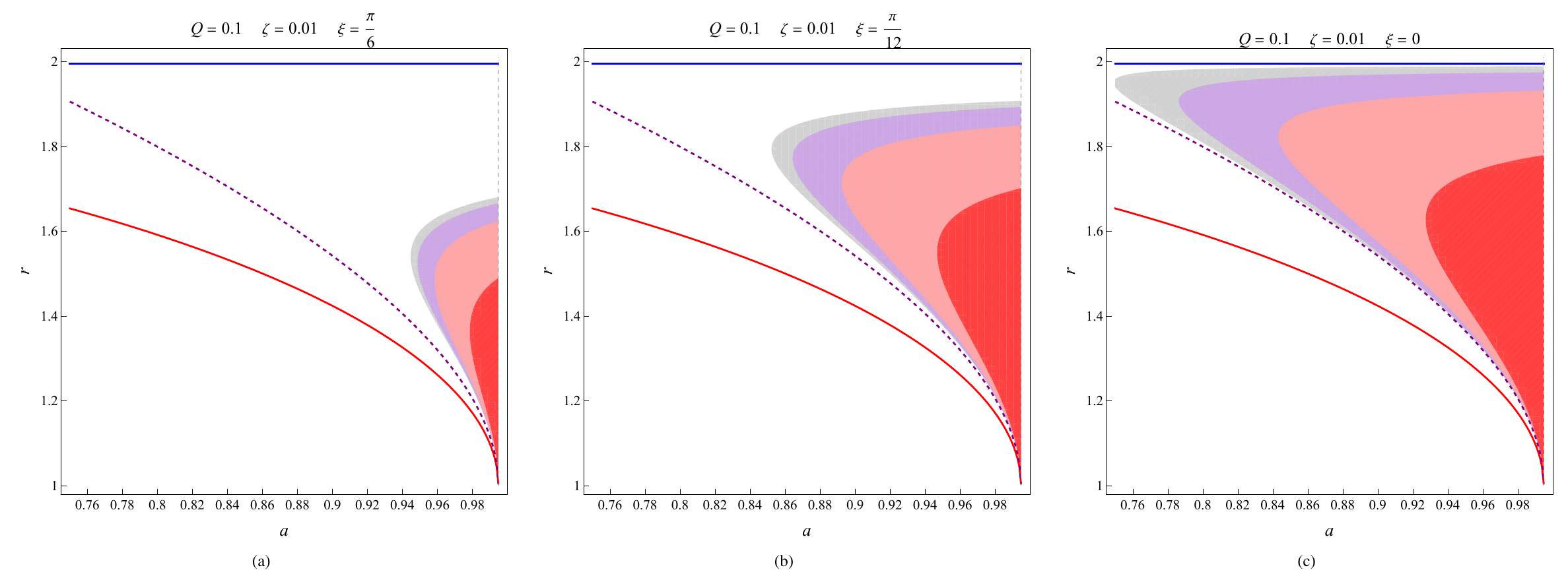}
    \caption{(Color online) Allowed energy extraction regions for different
    reconnection orientation angles $\xi$.}
    \label{fig:yxqxinltq}
\end{figure}

\begin{figure}[H]
    \centering
    \includegraphics[
        width=\linewidth,
        height=0.45\textheight,
        keepaspectratio
    ]{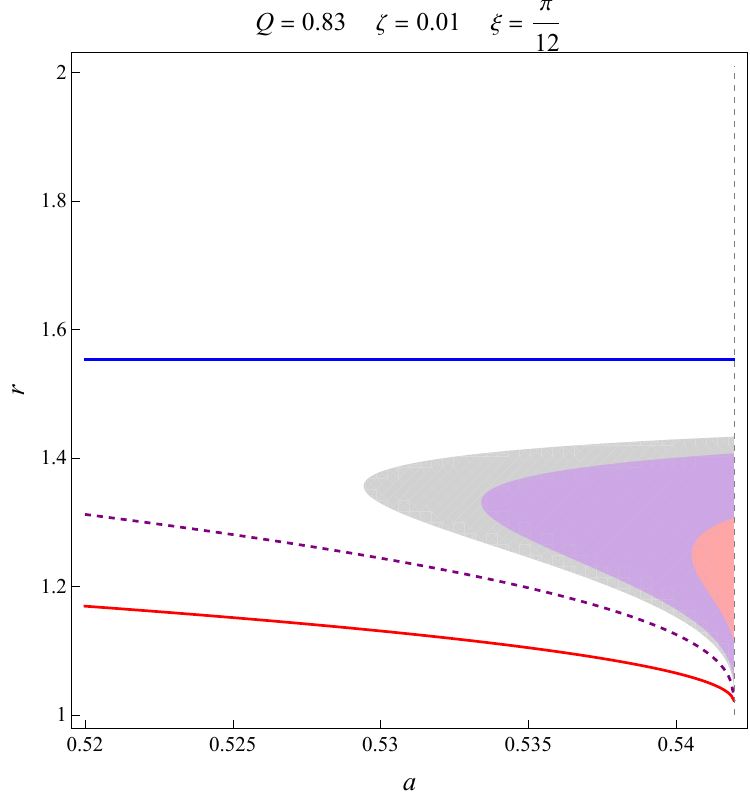}
    \caption{(Color online) Allowed energy extraction regions in the lower spin range.}
    \label{fig:cdzxnltq}
\end{figure}

The allowed energy extraction regions satisfying $e_-^\infty<0$ in the $(a,r)$ plane for different magnetic charge $Q$, coupling parameter $\zeta$, and orientation angle $\xi$ are displayed in Figs.~\ref{fig:xyqqnltq}, \ref{fig:yxqzetalntq}, and~\ref{fig:yxqxinltq}. Fig.~\ref{fig:cdzxnltq} further shows the allowed energy extraction region in the lower-spin range for a background with relatively large magnetic charge. In each panel, the red solid curve, blue solid curve, and purple dashed curve represent the event horizon, the ergoregion boundary, and the prograde circular photon orbit respectively, while the gray vertical dashed line on the right marks the maximum allowed spin for the corresponding parameters. The red, light red, purple, and gray shaded regions correspond to the magnetization parameters $\sigma=3,10,30,100$ respectively.

Fig.~\ref{fig:xyqqnltq} shows the allowed energy extraction regions for different backgrounds with different magnetic charges, with the coupling parameter fixed at $\zeta=0.01$ and the orientation angle fixed at $\xi=\pi/12$. For a given magnetic charge, as the magnetization parameter increases, the region satisfying the energy extraction condition extends toward lower spins and larger radii, indicating that a higher degree of magnetization is more favorable for the formation of negative energy outflows. As the magnetic charge increases, the allowed region shifts toward lower spins and smaller radii, while the maximum allowed spin of the black hole also decreases.

Fig.~\ref{fig:yxqzetalntq} shows the allowed energy extraction regions for different coupling parameters $\zeta$. For fixed magnetic charge and orientation angle, the allowed region shrinks overall as the coupling parameter increases, and its radial extent decreases, with the shrinkage being particularly pronounced for regions corresponding to lower magnetization parameters. At the same time, the maximum allowed spin of the black hole also decreases, further narrowing the spin range in which energy extraction can occur. On the other hand, increasing the magnetization parameter still causes the allowed region to expand toward lower spins and larger radii, indicating that a higher degree of magnetization helps maintain a broader parameter range for energy extraction even in the presence of stronger coupling. By contrast, as the coupling parameter increases, the allowed energy extraction region shrinks significantly, whereas the outer boundary of the ergoregion shifts only slightly inward, showing a much weaker change.

Fig.~\ref{fig:yxqxinltq} shows the allowed energy extraction regions for different orientation angles $\xi$. For fixed magnetic charge and coupling parameter, as the orientation angle decreases from $\pi/6$ to $0$, the allowed region extends significantly toward lower spins and larger radii. For $\xi=\pi/6$, energy extraction occurs mainly near the maximum allowed spin. By contrast, for $\xi=0$, the allowed region corresponding to larger magnetization parameters extends to the outer boundary of the ergoregion. 

To examine the lower-spin range in which magnetic reconnection energy extraction can occur, Fig.~\ref{fig:cdzxnltq} shows the allowed energy extraction region for $Q=0.83$, $\zeta=0.01$, and $\xi=\pi/12$. The results show that the allowed spin range for energy extraction is approximately $a=0.530$-$0.542$, indicating that magnetic reconnection energy extraction can also occur at relatively low spin parameters in backgrounds with large magnetic charge.

\subsection{Power and Efficiency of Energy Extraction in Circular Orbits}

To further quantify the rate at which energy is extracted from the black hole through magnetic reconnection, we investigate the corresponding energy extraction power. The power $P_{\rm extr}$ extracted from the black hole by the escaping plasma can be estimated \cite{Comisso2021,Wei2022,LiRegular2023} as 
\begin{equation}
   P_{\rm extr}=-e_-^\infty w_0 A_{\rm in}U_{\rm in}, 
   \label{eq:tqgl}
\end{equation}
where $w_0$ is the enthalpy density of the upstream plasma, $A_{\rm in}$ is the cross-sectional area of the inflow, and $U_{\rm in}$ is the reconnection inflow four-velocity. In this work, we consider collisionless reconnection and adopt $U_{\rm in}\simeq0.1$ \cite{Uzdensky2010}. For configurations in which prograde circular orbits exist within the ergoregion, we adopt the following approximation for the area: 
\begin{equation}
A_{\rm in}\sim r_{\rm E}^2-r_{\rm ph}^2,
   \label{eq:mjjs}
\end{equation}
where $r_{\rm E}$ is the ergoregion boundary and $r_{\rm ph}$ is the radius of the prograde photon orbit.

\begin{figure}[H]
    \centering
    \includegraphics[
        width=\linewidth,
        height=0.50\textheight,
        keepaspectratio
    ]{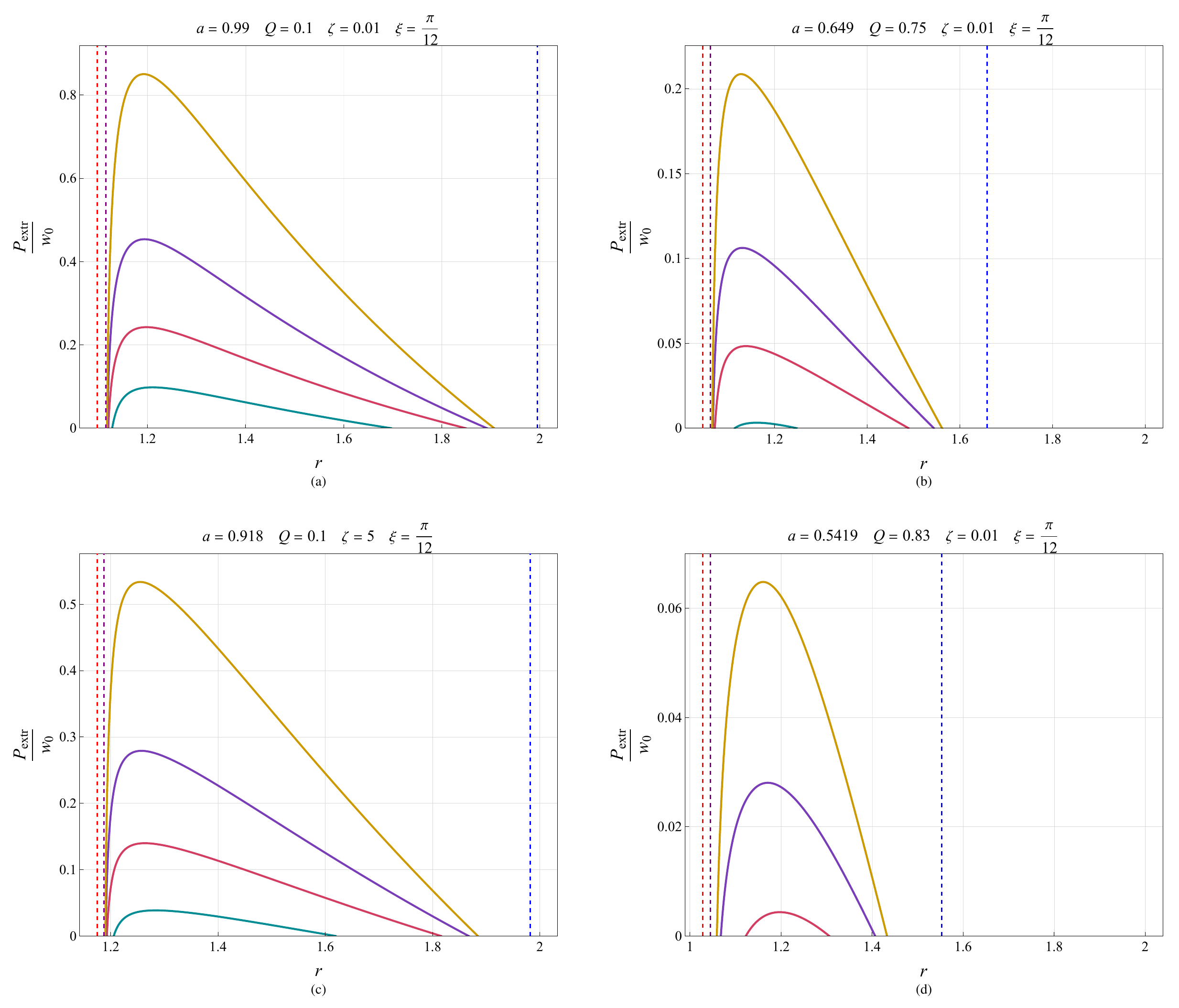}
    \caption{(Color online) Energy extraction power for different plasma magnetization parameters $\sigma$.}
    \label{fig:yxqtqgl}
\end{figure}

Fig.~\ref{fig:yxqtqgl} shows the variation of the normalized energy extraction power $P_{\mathrm{extr}}/w_0$ with the reconnection radius $r$ for different background parameters. The cyan, red, purple, and yellow curves correspond to the magnetization parameters $\sigma=3,10,30,100$ respectively. The red, purple, and blue vertical dashed lines indicate the event horizon, the prograde photon orbit, and the ergoregion boundary respectively. All power curves exhibit a single peak, first increasing and then decreasing. For the same background, increasing the magnetization parameter significantly enhances the peak power and broadens the radial range over which the power remains positive. 
The peak power and the radial extent of the positive power region differ considerably among different backgrounds. At the same magnetization parameter, a comparison of Figs.~\ref{fig:yxqtqgl}(a), (b), and (d) shows that, for a fixed coupling parameter $\zeta=0.01$, backgrounds with larger magnetic charge and lower spin correspond to lower peak power and a narrower radial range of positive power. A comparison of Figs.~\ref{fig:yxqtqgl}(a) and (c) further shows that, for a fixed magnetic charge $Q=0.1$, the peak power is also lower in the background with a larger coupling parameter. It should be noted that variations in the magnetic charge and coupling parameter also change the maximum allowed spin of the black hole. Therefore, different spin parameters are adopted in different panels. These comparisons reflect the differences in power resulting from the combined variations of the background parameters and the spin. In particular, Fig.~\ref{fig:yxqtqgl}(d) shows that, although energy extraction remains possible at relatively low spin parameters in the background with larger magnetic charge, the corresponding power is significantly reduced. The positive power associated with lower magnetization parameters is particularly weak.

We next investigate the efficiency of energy extraction through magnetic reconnection. The efficiency of energy extraction via magnetic reconnection is defined as
\cite{Comisso2021}
\begin{equation}
\eta=\frac{e_+^\infty}{e_+^\infty+e_-^\infty}.
   \label{eq:tqxl}
\end{equation}

\begin{figure}[H]
    \centering
    \includegraphics[
        width=\linewidth,
        height=0.50\textheight,
        keepaspectratio
    ]{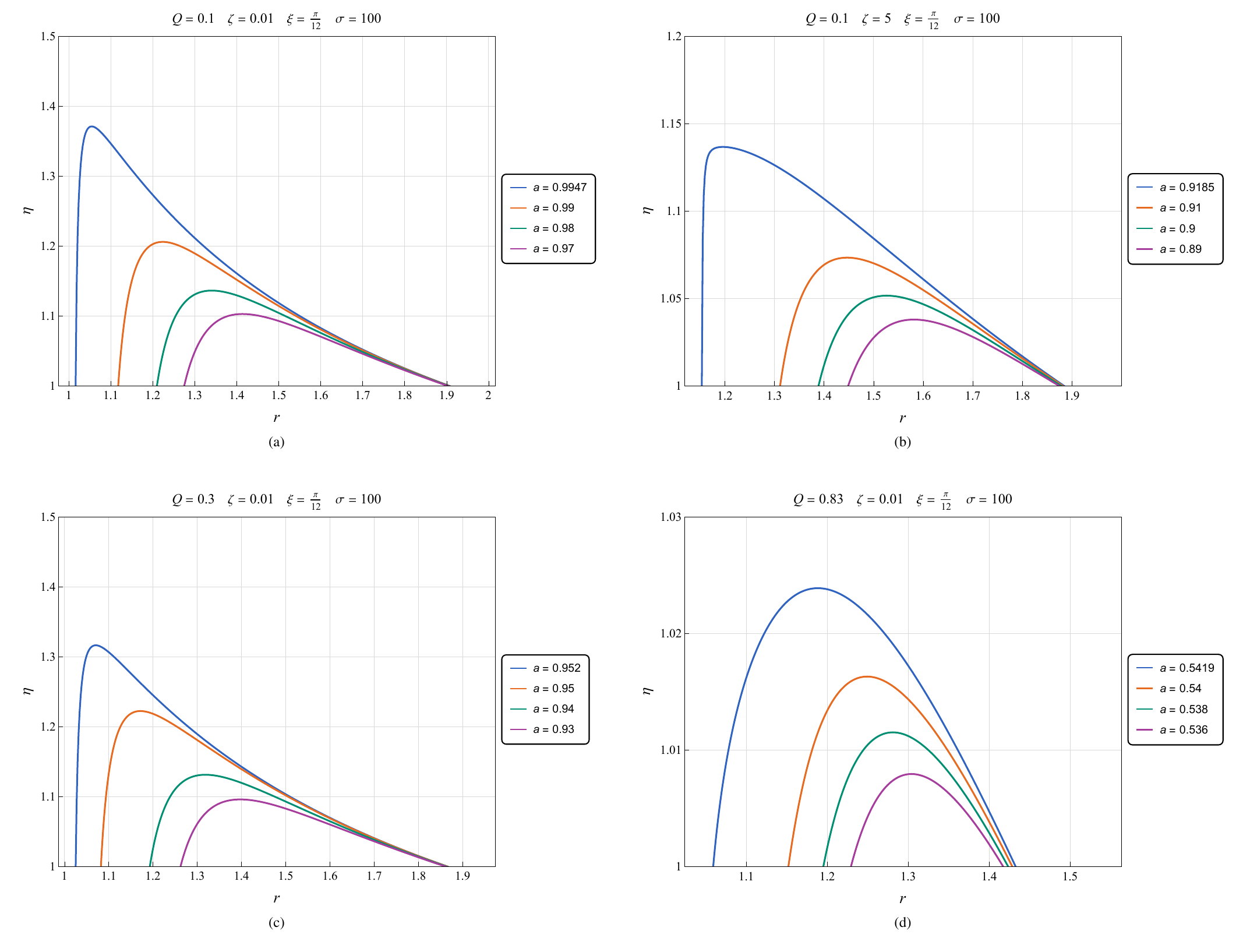}
    \caption{(Color online) Energy extraction efficiency for different
    spin parameters $a$.}
    \label{fig:yxqtqxl}
\end{figure}

Energy extraction from the black hole occurs when $\eta>1$. Fig.~\ref{fig:yxqtqxl} shows the variation of the energy extraction efficiency $\eta$ with the reconnection radius $r$ for different backgrounds, with $\sigma=100$ and $\xi=\pi/12$ fixed. All curves exhibit a single peak within the region where $\eta>1$, indicating the existence of an optimal reconnection radius at which the energy extraction efficiency reaches its maximum. For a given background, increasing the spin parameter significantly enhances the peak efficiency, shifts the peak position toward smaller radii, and extends the allowed radial range for energy extraction toward the black hole.

A further comparison among different backgrounds shows that the parameter combination corresponding to stronger coupling and lower spin in Fig.~\ref{fig:yxqtqxl}(b) produces a considerably lower peak efficiency than that in Fig.~\ref{fig:yxqtqxl}(a). For the panels with fixed $\zeta=0.01$, namely Figs.~\ref{fig:yxqtqxl}(a), (c), and (d), the peak efficiency of the highest spin curve generally decreases as the selected magnetic charge increases and the spin decreases. In particular, the efficiency in Fig.~\ref{fig:yxqtqxl}(d) is only slightly above unity, indicating that although energy extraction remains possible in this background, the efficiency gain is relatively small. Since different spin parameters are adopted for different backgrounds, these comparisons reflect the efficiency differences arising from the combined variations of the background parameters and the spin.

\section{Energy Extraction in the Plunging Region} \label{SecEEP}

\subsection{Parameter Space for Allowed Energy Extraction in the Plunging Region}

In the preceding section, we considered the case in which the plasma moves along circular orbits in the equatorial plane. We now further consider the case in which the plasma moves along prograde circular orbits outside the innermost stable circular orbit (ISCO) and plunges toward the black hole inside the ISCO \cite{Wilkins2020}. Let $r_I$ denote the ISCO radius. The plunging region outside the event horizon is defined as $r_H<r<r_I$. Since the plunging plasma possesses a radial velocity, both radial and azimuthal motions must be taken into account. We continue to adopt the ZAMO frame, in which the background four velocity is expressed as \cite{Shen2024}
\begin{equation}
\hat U^{(\mu)}
=\hat\gamma_s
\left\{1,\hat v_s^{(r)},0,\hat v_s^{(\phi)}\right\}
=
\left\{
\frac{E-\omega^\phi L}{\alpha},
\sqrt{g_{rr}}\,U^r,
0,
\frac{L}{\sqrt{g_{\phi\phi}}}
\right\},
\label{eq:ssd}
\end{equation}
where $U^r=dr/d\tau$.

For equatorial circular orbits, the specific energy $E_K(r)$ and specific angular momentum $L_K(r)$ are given by \cite{Bardeen1972,Harko2009}
\begin{equation}
\begin{aligned}
E_K(r)
&=-\frac{g_{tt}+g_{t\phi}\Omega_K}
{\sqrt{-g_{tt}-2g_{t\phi}\Omega_K
-g_{\phi\phi}\Omega_K^2}},\\
L_K(r)
&=\frac{g_{t\phi}+g_{\phi\phi}\Omega_K}
{\sqrt{-g_{tt}-2g_{t\phi}\Omega_K
-g_{\phi\phi}\Omega_K^2}}.
\end{aligned}
\label{eq:keplerEL}
\end{equation}
where $\Omega_K$ is the Keplerian angular velocity defined in Eq.~\ref{eq:kpljsd}.

During the plunging motion, the plasma retains the specific energy and angular momentum of the circular orbit at the ISCO, given by
\begin{equation}
E_I=E_K(r_I),\qquad
L_I=L_K(r_I).
\label{eq:iscoshl}
\end{equation}
Here, $r_I$ denotes the ISCO radius, while $E_I$ and $L_I$ are the conserved specific energy and angular momentum during the plunging motion respectively.

For timelike particles, Eq.~\ref{eq:jxs} is evaluated with $\varepsilon=1$, and the ISCO satisfies
\begin{equation}
\mathcal R(r_I)=0,\qquad
\left.\partial_r\mathcal R\right|_{r_I}=0,\qquad
\left.\partial_r^2\mathcal R\right|_{r_I}=0.
\end{equation}

Substituting Eq.~\ref{eq:iscoshl} into Eq.~\ref{eq:jxs}, we obtain
\begin{equation}
    U^r= \frac{\mathrm{d}r}{\mathrm{d}\tau}=-\frac{\sqrt{\mathcal R(r)}}{r^2}，
\end{equation}
where $\mathcal R(r)$ is evaluated with $E=E_I$, $L=L_I$ and $\varepsilon=1$. The negative sign indicates inward motion toward the black hole.

The velocity components $\hat v_s^{(r)}$ and $\hat v_s^{(\phi)}$ can be determined from Eq.~\ref{eq:ssd}. The corresponding total velocity and Lorentz factor are
\begin{equation}
\hat v_s
=\sqrt{\bigl(\hat v_s^{(r)}\bigr)^2+
       \bigl(\hat v_s^{(\phi)}\bigr)^2},
\qquad
\hat\gamma_s=(1-\hat v_s^2)^{-1/2}.
\end{equation}

Using the same hot fluid approximation as in the circular orbit section, the energies at infinity per unit enthalpy of the two plasma outflows in the plunging background can be written as \cite{Chen2024,Shen2025}
\begin{equation}
e_\pm^\infty
=\alpha\hat\gamma_s\gamma_{\rm out}
\Bigg[1+\beta^\phi\hat v_s^{(\phi)}\pm v_{\rm out}
\left(\hat v_s+\beta^\phi\frac{\hat v_s^{(\phi)}}{\hat v_s}\right)\cos\xi \mp v_{\rm out}\beta^\phi\frac{\hat v_s^{(r)}}{\hat\gamma_s\hat v_s}\sin\xi
\Bigg]-\frac{\alpha}{4\hat\gamma_s\gamma_{\rm out}\left(1\pm\hat v_s v_{\rm out}\cos\xi\right)},
\end{equation}
where the outflow velocity $v_{\rm out}$ and its Lorentz factor $\gamma_{\rm out}$ are given by
\begin{equation}
v_{\rm out}=\sqrt{\frac{\sigma}{1+\sigma}},
\qquad
\gamma_{\rm out}=\sqrt{1+\sigma}.
\end{equation}

\begin{figure}[H]
    \centering
    \includegraphics[
        width=\linewidth,
        height=0.45\textheight,
        keepaspectratio
    ]{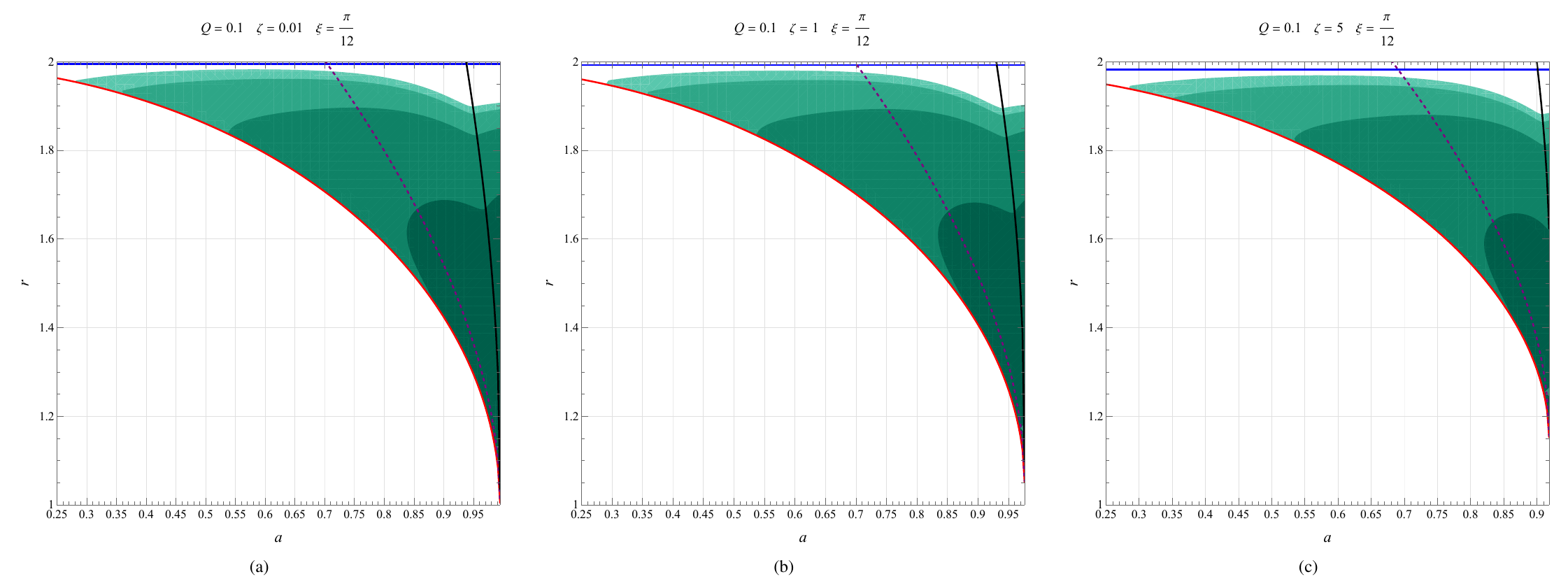}
    \caption{(Color online) Allowed energy extraction regions in the
    plunging region for different coupling parameters $\zeta$.}
    \label{fig:pzetanltq}
\end{figure}

\begin{figure}[H]
    \centering
    \includegraphics[
        width=\linewidth,
        height=0.45\textheight,
        keepaspectratio
    ]{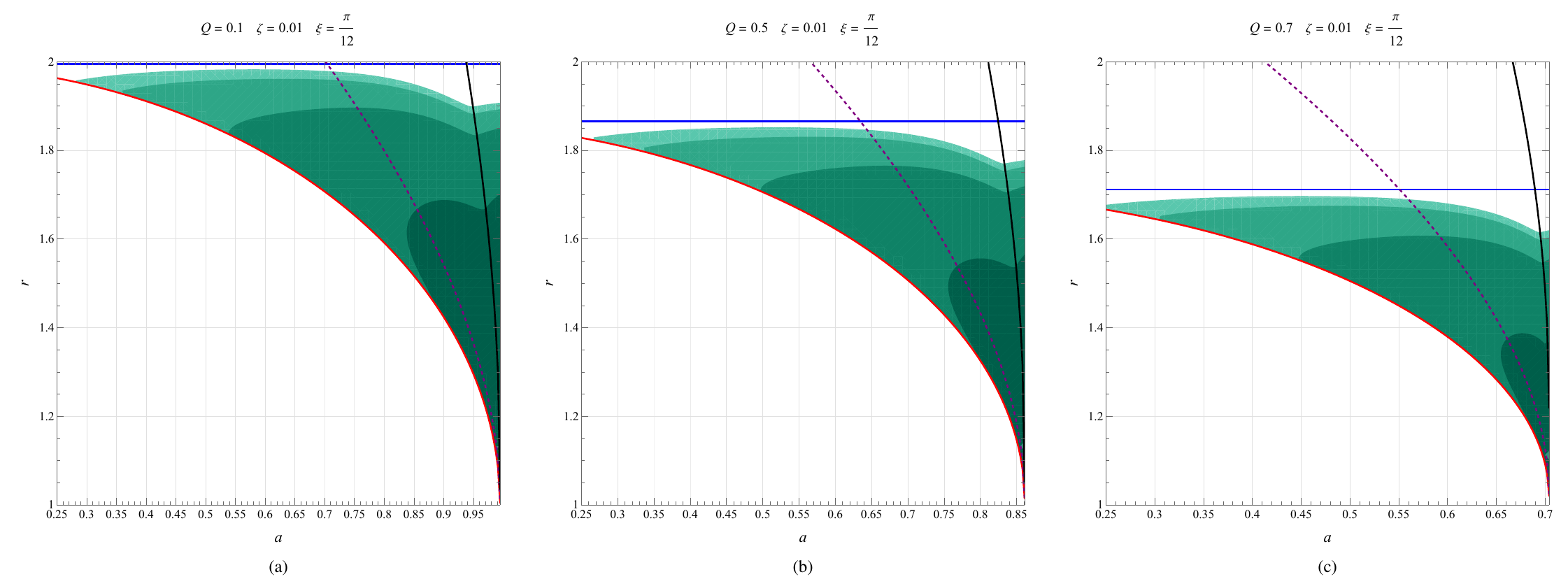}
    \caption{(Color online) Allowed energy extraction regions in the
    plunging region for different magnetic charges $Q$.}
    \label{fig:pqnltq}
\end{figure}

\begin{figure}[H]
    \centering
    \includegraphics[
        width=\linewidth,
        height=0.45\textheight,
        keepaspectratio
    ]{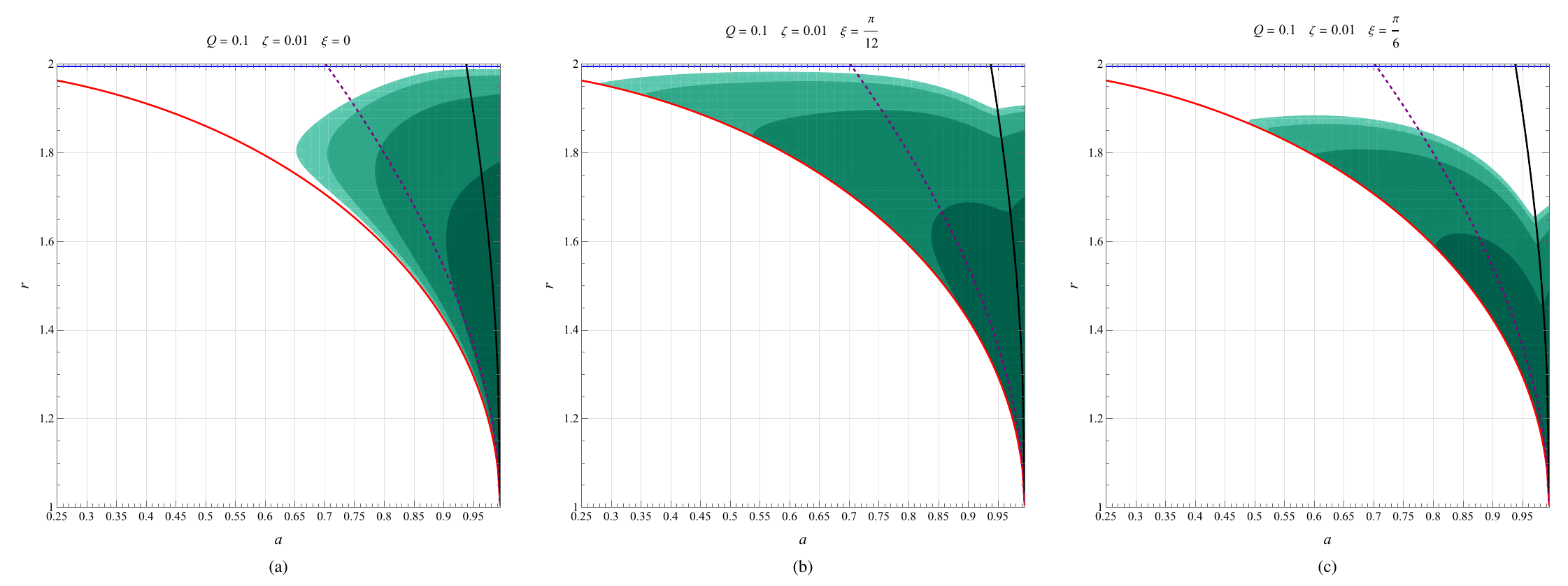}
    \caption{(Color online) Allowed energy extraction regions in the
    plunging region for different reconnection orientation
    angles $\xi$.}
    \label{fig:pxinltq}
\end{figure}

To investigate the influence of plunging motion on energy extraction via magnetic reconnection, Figs.~\ref{fig:pzetanltq},~\ref{fig:pqnltq}, and~\ref{fig:pxinltq} show the allowed regions satisfying $e_-^\infty<0$ in the $(a,r)$ plane for different values of the coupling parameter $\zeta$, magnetic charge $Q$, and reconnection orientation angle $\xi$ respectively. The green shaded regions, from light to dark, correspond to magnetization parameters $\sigma=100,30,10,3$ respectively, with the different regions overlapping. The red solid curve, blue solid curve, purple dashed curve, and black solid curve represent the event horizon, the ergoregion boundary, the prograde photon orbit, and the innermost stable circular orbit (ISCO) respectively. The region between the event horizon and the ISCO corresponds to the plunging region, whereas the circular orbit region is adopted outside the ISCO.

Fig.~\ref{fig:pzetanltq} shows the allowed energy extraction regions for different coupling parameters $\zeta$, with the magnetic charge fixed at $Q=0.1$ and the orientation angle fixed at $\xi=\pi/12$. As the magnetization parameter $\sigma$ increases, the allowed region extends toward lower spins and larger radii. In the lower spin range on the left side, the allowed regions corresponding to higher magnetization parameters are located close to the event horizon and have a relatively narrow radial extent. As the spin parameter $a$ increases, the radial extent of these regions gradually broadens. In comparison, variations in the coupling parameter do not significantly change the overall structure of the allowed regions. As $\zeta$ increases, the maximum allowed spin decreases, causing the high spin end of the allowed region to shift toward smaller values of $a$, while the outer boundary of the ergoregion moves only slightly inward. Compared with the results for the circular orbit region shown in Fig.~\ref{fig:yxqzetalntq}, the allowed energy extraction regions in the plunging region exhibit weaker sensitivity to variations in the coupling parameter. Within the common spin range of the different panels, the allowed regions corresponding to different magnetization parameters retain similar overall structures as $\zeta$ increases. The more noticeable changes mainly occur near the high spin end. This difference indicates that the influence of the coupling parameter on the allowed energy extraction regions depends on the state of plasma motion.

Compared with the variations caused by changing the coupling parameter in Fig.~\ref{fig:pzetanltq}, the changes in the positions of the allowed regions induced by variations in the magnetic charge are more pronounced in Fig.~\ref{fig:pqnltq}. As the magnetic charge $Q$ increases, the outer boundary of the ergoregion moves toward smaller radii, and the allowed energy extraction regions are consequently distributed at smaller radii. Meanwhile, the maximum allowed spin of the black hole decreases, shifting the high spin end of the allowed
regions toward smaller values of $a$. For different magnetic charge backgrounds, increasing the magnetization parameter $\sigma$ expands both the spin and radial ranges of the allowed regions. The regions corresponding to lower magnetization parameters are mainly concentrated in the higher spin ranges of their respective backgrounds.

Fig.~\ref{fig:pxinltq} show the allowed energy extraction regions for different reconnection orientation angles $\xi$ with $Q=0.1$ and $\zeta=0.01$ fixed. Since the background parameters of the black holes remain unchanged, the event horizon, outer boundary of the ergoregion, prograde photon orbit, and ISCO are identical in all panels, whereas the allowed regions change significantly with the orientation angle.

As the parameter $\xi$ increases from $0$ to $\pi/12$, the allowed region extends significantly toward lower spins. When $\xi$ further increases to $\pi/6$, the lower spin boundary shifts toward larger values of $a$, while the upper radial boundary also decreases. Therefore, among the three orientation angles shown, $\xi=\pi/12$ corresponds to an allowed region extending to the lowest spin values. However, at higher spins, the allowed region for $\xi=0$ can extend to larger radii, indicating that the allowed regions for different orientation angles are not simply nested within one another. This result shows that the angular dependence of the allowed regions in the plunging region cannot be summarized by stating that smaller orientation angles are always more favorable. Furthermore, increasing the magnetization parameter in all three panels expands the allowed regions, allowing negative energy outflows to form over broader ranges of spin and radius.

\subsection{Power and Efficiency of Energy Extraction in the Plunging Region}

\begin{figure}[H]
    \centering
    \includegraphics[
        width=\linewidth,
        height=0.45\textheight,
        keepaspectratio
    ]{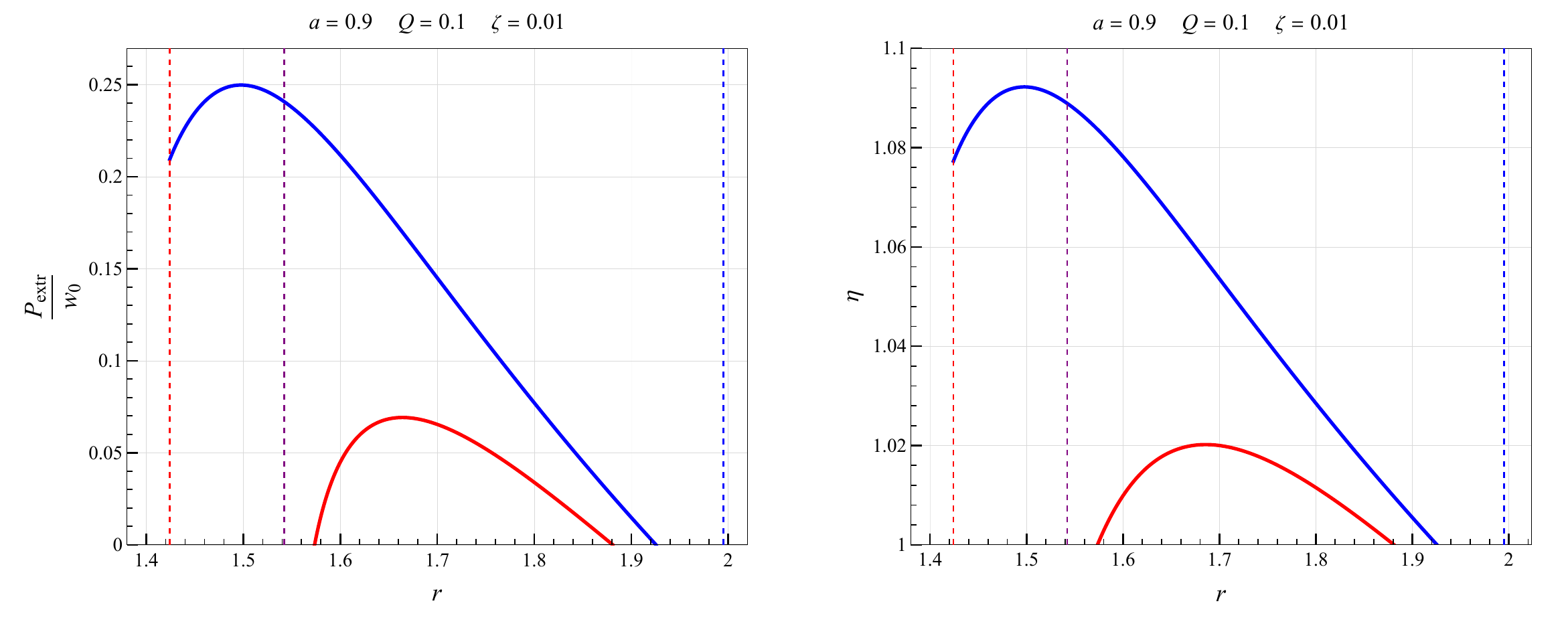}
    \caption{(Color online) Comparison of the power and efficiency
    in the plunging region and the circular orbit region.}
    \label{fig:pydb}
\end{figure}

Based on the allowed energy extraction regions discussed above, Fig.~\ref{fig:pydb} further compares the energy extraction power and efficiency of the plunging and circular orbit regions. We fix $a=0.9$, $Q=0.1$, $\zeta=0.01$, $\sigma=100$ and $\xi=\pi/12$. The blue and red solid curves correspond to the plunging plasma and circular orbit plasma respectively, while the red, purple, and blue vertical dashed lines represent the event horizon, prograde photon orbit, and ergoregion boundary respectively. Both the power and efficiency in the two regions first increase and then decrease with the radius, but the peak values of the plunging region are significantly higher.

\section{Conclusions} \label{SecCOn}

In this work, we have investigated energy extraction via magnetic reconnection in the background of rotating Einstein-Yang-Mills black holes with nonminimal coupling within the Comisso-Asenjo mechanism. Both circular orbit motion and plunging motion from the innermost stable circular orbit have been considered. We first analyzed the event horizon, the outer boundary of the equatorial ergoregion, and the circular photon orbit. We then investigate how magnetic charge \(Q\), coupling parameter \(\zeta\), magnetization parameter \(\sigma\), and orientation angle \(\xi\) affect outflow energy, allowed energy extraction regions, power, and efficiency.

In the circular orbit region, the effect of the magnetic charge \(Q\) on the energy at infinity \(e_-^\infty\) of the decelerated plasma depends on the reconnection radius. Increasing \(Q\) raises \(e_-^\infty\) on one side of the reversal boundary and lowers it on the other. We locate this boundary by identifying where \(\partial e_-^\infty/\partial Q\) changes sign. Within the parameter range considered, increasing the coupling parameter \(\zeta\) shifts the reversal boundary toward larger radii. This shift broadens the range over which increasing \(Q\) reduces \(e_-^\infty\).

We compare the allowed energy extraction regions for circular and plunging motion. Increasing the magnetization parameter \(\sigma\) expands the allowed region in both cases.Decreasing the orientation angle \(\xi\) expands the region for circular motion, but the region for plunging motion does not expand monotonically. For circular motion, increasing the coupling parameter \(\zeta\) markedly shrinks the allowed region, while the outer boundary of the equatorial ergoregion shifts only slightly inward.For plunging motion, the overall shape of the allowed region remains similar across different values of \(\zeta\) within the common spin range. However, the maximum allowed spin and the boundary at the high-spin end both shift. Energy extraction is also possible at lower spins for plunging motion than for circular motion. For the parameter sets considered, energy extraction in the plunging region achieves both higher power and higher efficiency than in the circular orbit region.

These results indicate that the effects of magnetic charge and nonminimal coupling on energy extraction through magnetic reconnection depend not only on the background parameters themselves, but also strongly on the reconnection radius and the state of plasma motion.

\end{document}